\documentclass[final]{IEEEtran}
\usepackage{cite}
\usepackage{threeparttable}
\usepackage{graphicx}
\usepackage{picinpar}
\usepackage[cmex10]{amsmath}
\usepackage{amsmath,amsfonts,amssymb}
\usepackage[linesnumbered,ruled,lined,boxed]{algorithm2e}
\usepackage{subfigure}
\usepackage{threeparttable}
\usepackage{caption}
\usepackage{booktabs}
\usepackage{multirow}
\usepackage{diagbox}
\usepackage{etoolbox}
\usepackage{makecell}
\usepackage{amssymb}
\usepackage{mathrsfs}
\usepackage{color}
\definecolor{purple}{RGB}{160,32,240}
\definecolor{darkred}{RGB}{255,0,255}
\usepackage{stfloats}
\usepackage{bm}

\begin{document}

\newtheorem{lemma}{Lemma}
\newtheorem{corol}{Corollary}
\newtheorem{theorem}{Theorem}
\newtheorem{proposition}{Proposition}
\newtheorem{definition}{Definition}
\newcommand{\e}{\begin{equation}}
\newcommand{\ee}{\end{equation}}
\newcommand{\eqn}{\begin{eqnarray}}
\newcommand{\eeqn}{\end{eqnarray}}

\title{\Large Compressive Sensing Based Massive Access for IoT Relying on Media Modulation Aided Machine Type Communications}
\author{Li Qiao, Jun Zhang, Zhen Gao, Sheng Chen,~\IEEEmembership{Fellow,~IEEE,} and Lajos Hanzo,~\IEEEmembership{Fellow,~IEEE}%
 \vspace*{-3.0mm}
\thanks{L. Qiao, J. Zhang and Z. Gao are with the Advanced Research Institute of Multidisciplinary Science and School of Information and Electronics, Beijing Institute of Technology, Beijing 100081, China.}
 \thanks{S. Chen is with the School of Electronics and Computer Science, University
of Southampton, Southampton SO17 1BJ, U.K., and also with King Abdulaziz
University, Jeddah 21589, Saudi Arabia.}  %
 \thanks{L. Hanzo is with the School of Electronics and Computer Science, University
of Southampton, Southampton SO17 1BJ, U.K.}  %
}

\maketitle

\begin{abstract}
A fundamental challenge of the large-scale Internet-of-Things lies in how to support massive machine-type communications (mMTC). This letter proposes a media modulation based mMTC solution for increasing the throughput, where a massive multi-input multi-output based base station (BS) is used for enhancing the detection performance. For such a mMTC scenario, the reliable active device detection and data decoding pose a serious challenge. By leveraging the sparsity of the uplink access signals of mMTC received at the BS, a compressive sensing based massive access solution is proposed for tackling this challenge. Specifically, we propose a structured orthogonal matching pursuit algorithm for detecting the active devices, whereby the block-sparsity of the uplink access signals exhibited across the successive time slots and the structured sparsity of media modulated symbols are exploited for enhancing the detection performance. Moreover, a successive interference cancellation based structured subspace pursuit algorithm is conceived for data demodulation of the active devices, whereby the structured sparsity of media modulation based symbols found in each time slot is exploited for improving the detection performance. Finally, our simulation results verify the superiority of the proposed scheme over state-of-the-art solutions.
\end{abstract}

\vspace*{-1mm}
\begin{IEEEkeywords}
Internet-of-Things, massive machine type communications, massive access, massive multi-input multi-output, media modulation, compressive sensing.
\end{IEEEkeywords}
 \vspace*{-2.0mm}

\IEEEpeerreviewmaketitle

\section{Introduction}

\IEEEPARstart{T}{}he emerging paradigm of massive machine-type communications (mMTC) is identified as an indispensable component for enabling the massive access of machine-type devices (MTDs) in the emerging Internet-of-Things (IoT) \cite{overview1}. In stark contrast to conventional human-centric mobile communications, mMTC focuses on uplink-oriented communications serving massive MTDs and exhibits sporadic tele-traffic requiring low-latency and high-reliability massive access \cite{overview1}.

The conventional grant-based access approach relies on complex time and frequency-domain resource allocation before data transmission, which would impose prohibitive signaling overhead and latency on massive mMTC \cite{overview1}. To support low-power MTDs at low latency, the emerging grant-free approach has attracted significant attention for massive access, since it simplifies the access procedure by directly delivering data without scheduling \cite{BWang1,YangDU1,Profshim2,BWang2,YangDU2,TWOLEVEL}. Specifically, by exploiting the block-sparsity of mMTC, the authors of \cite{BWang1} and \cite{YangDU1} proposed compressive sensing (CS) solutions for joint active device and data detection, while a maximum {\it a posteriori} probability based scheme was proposed in \cite{Profshim2} for improving performance. Furthermore, MTDs having slow-varying activity tend to exhibit partially block sparsity, hence a modified orthogonal matching pursuit solution was conceived in \cite{BWang2}, beside a modified subspace pursuit algorithm was proposed in \cite{YangDU2}. It was shown that the previous detected results can be exploited for enhancing the following detection. However, the contributions \cite{BWang1,YangDU1,Profshim2,BWang2,YangDU2} only consider single-antenna configurations at both the MTDs and the BS. To achieve higher efficiency and more reliable detection, multi-antenna at MTDs using spatial modulation (SM) and massive multi-input multi-output (mMIMO) were considered in \cite{Gao,TWOLEVEL} at the BS, where a two-level sparse structure based CS (TLSSCS) detector and a structured CS detector was proposed in \cite{TWOLEVEL} and \cite{Gao}, respectively. However, the increased data rate of SM by one bit requires doubling the number of antennas \cite{SM1,SM2}, which violates the low-cost requirement of MTDs. To improve the uplink (UL) throughput at a low cost and power-consumption, authors of \cite{MBMMUD1,MBMMUD2} proposed to employ media modulation at the MTDs, where an iterative interference cancellation detector and a CS detector was employed for multi-user detection in \cite{MBMMUD1} and \cite{MBMMUD2}, respectively. However, they have not considered the active user detection (AUD). {\color{black} To sum up, we provide a brief comparison of the related literatures in Table I.}

Against this background, we propose to adopt media modulation at the MTDs for improving the UL throughput and to employ a mMIMO scheme at the BS. Moreover, a CS-based active device and data detection solution is proposed by exploiting both the sporadic traffic and the block-sparsity of mMTC as well as the structured sparsity of media modulated symbols. Specifically, we first propose a structured orthogonal matching pursuit (StrOMP) algorithm for AUD, where the block-sparsity of UL access signals across the successive time slots and the structured sparsity of media modulated symbols are exploited. Additionally, a successive interference cancellation based structured subspace pursuit (SIC-SSP) algorithm is proposed for demodulating the detected active MTDs, where the structured sparsity of media modulated symbols in each time slot is exploited for enhancing the decoding performance.
{\color{black}
Note that the proposed CS-based StrOMP and SIC-SSP algorithms belong to the greedy algorithms. Due to the computational benefit and near-optimal performance, greedy algorithms have been popularly used in the mMTC scenarios \cite{Profshim2,BWang1,YangDU1,BWang2,YangDU2,TWOLEVEL,CSreview}.} Finally, our simulation results verify the superiority of the proposed scheme over cutting-edge benchmarks.

{\textit {Notation}: Boldface lower and upper-case symbols denote column vectors and matrices, respectively.
For a matrix ${\bf A}$, ${\bf A}^T$, ${\bf A}^H$, ${\bf A}^\dagger$, ${\left\| {\bf{A}} \right\|_F}$, ${\bf{A}}_{[m,n]}$ denote the transpose, Hermitian transpose, pseudo-inverse, Frobenius norm, the $m$-th row and $n$-th column element of ${\bf{A}}$, respectively. ${\bf{A}}_{[\Omega,:]}$ (${\bf{A}}_{[:,\Omega]}$) is the sub-matrix containing the rows (columns) of ${\bf{A}}$ indexed in the ordered set $\Omega$. ${\bf{A}}_{[\Omega,m]}$ is the $m$-th column of ${\bf{A}}_{[\Omega,:]}$. For a vector ${\bf x}$, ${\left\| {\bf x} \right\|_p}$, $[{\bf x}]_{m}$, $[{\bf x}]_{m:n}$ and $[{\bf x}]_{\Omega}$ are the ${l_p}$ norm, $m$-th element, $m$-th to $n$-th elements, and entries indexed in the ordered set $\Omega$ of ${\bf x}$, respectively. For an ordered set $\Gamma$ and its subset $\Omega$, $|\Gamma|_c$, $\Gamma[m]$, and $\Gamma\setminus \Omega$ are the cardinality of $\Gamma$, $m$-th element of $\Gamma$, and complement of subset $\Omega$ in $\Gamma$, respectively. $[K]$ is the set $\{1,2,...,K\}$.

\begin{table}[!t]
\centering
\captionsetup{font = {normalsize, color = {black}}, labelsep = period} 
\color{black}\caption*{Table I: A brief comparison of the related literatures}
\begin{tabular}{|c|c|c|c|c|c|c|}
\Xhline{1.2pt}
\multicolumn{2}{|c|}{\diagbox{Contents}{Literatures}} & [2]-[6] & [7] & [8] & [11] & [12]\\%
\Xhline{1.2pt}
\multirow{2}*{BS}
&Single antenna &\checkmark& & & & \\
\cline{2-7}
&mMIMO & & \checkmark& \checkmark & \checkmark &\checkmark\\
\Xhline{1.2pt}
\multirow{3}*{MTDs}
&Single antenna&\checkmark& & & &\\
\cline{2-7}
&SM & & \checkmark& \checkmark &  &\\
\cline{2-7}
&Media modulation &  & & & \checkmark& \checkmark\\
\Xhline{1.2pt}
\multicolumn{2}{|c|}{AUD}&\checkmark & \checkmark&  &  &\\
\Xhline{1.2pt}
\multicolumn{2}{|c|}{Data detection}&\checkmark & \checkmark& \checkmark & \checkmark &\checkmark\\
\Xhline{1.2pt}
\end{tabular}
\end{table}

\vspace{-2mm}
\section{System Model}

We first introduce the proposed media modulation based mMTC scheme and then focus on our massive access technique relying on joint active device and data detection at the BS.

\vspace{-3.5mm}
\subsection{Proposed Media Modulation Based mMTC Scheme}

As illustrated in Fig. \ref{fig:MTC-MBM-joint}, we propose that all $K$ MTDs adopt media modulation for enhanced UL throughput and the BS employs mMIMO using $N_r$ receive antenna elements for reliable massive access. In the UL, each symbol consists of the conventional modulated symbol and of the media modulated symbol, and each MTD relies on a single conventional antenna and $M_r$ extra radio frequency (RF) mirrors \cite{MBMMUD1,MBMMUD2,MBM1,MBM2,MBM3}. By adjusting the binary on/off status of the $M_r$ RF mirrors, we have $N_t=2^{M_r}$ mirror activation patterns (MAPs), and the media modulated symbol is obtained by mapping ${\rm log_2}{(N_t)}=M_r$ bits to one of the $N_t$ MAPs. Therefore, if the conventional $M$-QAM symbol is adopted, the overall UL throughput of an MTD is
$\eta = M_r +{ \it{\rm log_2}} M$ bit per channel use (bpcu). {\color{black}By contrast, to realize the extra bits, SM technique with an RF chain and multiple transmit antennas will select one of the transmit antennas for UL transmission \cite{SM1,SM2}. Hence, to achieve the same extra throughput, media modulation requires a single UL transmit antenna and a linearly increasing number of RF mirrors, but SM requires an exponentially increasing number of antennas \cite{MBMMUD1,MBMMUD2,SM1,SM2,MBM1,MBM2,MBM3}. Clearly, media modulation is more attractive for mMTC owing to its increased UL throughput at a negligible power consumption and hardware cost \cite{MBM1,MBM2,MBM3}.} Moreover, using a mMIMO UL receiver is the most compelling technique. By leveraging the substantial diversity gain gleaned from hundreds of antennas, the mMIMO BS is expected to achieve high-reliability UL multi-user detection, in the context of mMTC. By integrating the complementary benefits of media modulation at the MTDs and mMIMO reception at the BS into mMTC, we arrive at an excellent solution.

\vspace{-3.5mm}
\subsection{Massive Access of the Proposed mMTC Scheme}

As shown in Fig. \ref{fig:MTC-MBM-joint}, we assume that the activity patterns of the $K$ MTDs remain unchanged in a frame, which consists of $J$ successive time slots. Hence we only focus our attention on the massive access for a given frame. Specifically, the signal received at the BS in the $j$-th $(\forall j\in [J])$ time slot, denoted by ${\bf{y}}^j\in\mathbb{C}^{N_r \times 1}$, can be expressed as
\begin{equation}\label{eq:system}
\begin{array}{l}
{\bf y}^j=\sum\limits_{k = 1}^K a_k g_k^j {\bf H}_k {\bf d}_k^j+{\bf w}^j=\sum\limits_{k = 1}^K {\bf H}_k {\bf x}_k^j+{\bf w}^j
={\bf H} \widetilde{{\bf x}}^j +  {\bf w}^j,
\end{array}
\end{equation}
where the activity indicator $a_k$ is set to one (zero) if the $k$-th MTD is active (inactive), while ${g_k^{j}}\in\mathbb{C}$, ${\bf d}_k^j\in\mathbb{C}^{N_t \times 1}$, and ${\bf x}_k^j=a_k g_k^j {\bf d}_k^j \in\mathbb{C}^{N_t \times 1}$ are the conventional modulated symbol, media modulated symbol, and equivalent UL access symbol of the $k$-th MTD in the $j$-th time slot, respectively. Furthermore, ${\bf H}_k\in\mathbb{C}^{N_r\times N_t}$ is the multi-input multi-output (MIMO) channel matrix associated with the $k$-th MTD, ${\bf w}^j \in \mathbb{C}^{N_r \times 1}$ is the noise with elements obeying the independent and identically distributed (i.i.d.) complex Gaussian distribution ${\cal C}{\cal N}( {0,\sigma_w^2} )$, while ${{\bf{ H}}} = [{\bf H} _1, {\bf H} _2,...,{\bf H} _K] \in \mathbb{C}^{N_r \times (K N_t)}$ and $\widetilde{\bf x}^j = [({\bf x}^j _1)^T,({\bf x}^j _2)^T,...,({\bf x}^j_K)^T]^T \in \mathbb{C}^{(K N_t) \times 1}$ are the aggregate MIMO channel matrix and UL access signal in the $j$-th time slot, respectively.
\begin{figure}
     \centering
     \includegraphics[width=8.7cm,height=5.5cm, keepaspectratio]%
     {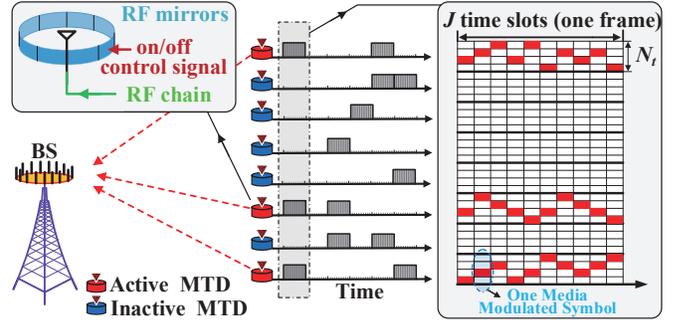}
     \captionsetup{font={footnotesize}, name={Fig.},labelsep=period}
     \caption{Proposed media modulation based mMTC scheme, where the UL access signal exhibits block-sparsity in a frame and structured sparsity in each time slot.}
     \label{fig:MTC-MBM-joint}
     \vspace*{-5mm}
\end{figure}

Note that for any ${\bf d}_k^j$ given $\forall j\in[J]$ and $\forall k\in[K]$, only one of its entries is one and the others are all zeros, i.e.,
\begin{equation}\label{eq:d}
\begin{array}{l}
{\rm supp}\{{\bf d}_k^j\}\in [N_t],~~\parallel\!{{\bf d}_k^j}\!\parallel_0=1,~~\parallel\!{{\bf d}_k^j}\!\parallel_2=1,
\end{array}
\end{equation}
where ${\rm supp\{\cdot\}}$ is the support set of its argument. Furthermore, we consider the Rayleigh MIMO channel model, hence the elements in ${\bf H}_k$ for $\forall k\!\in\![K]$ follow the i.i.d. complex Gaussian distribution ${\cal C}{\cal N}(0,1)$. We assume that the channels remain time-invariant for a relatively long period in typical IoT scenarios, hence $\{{\bf H}_k\}_{k=1}^K$ can be accurately estimated at the BS via periodic updates.

\section{Proposed CS-Based Massive Access Solution}%

In typical IoT scenarios, the MTDs generate sporadic tele-traffic \cite{Profshim2,BWang1,YangDU1,BWang2,YangDU2,TWOLEVEL}, which indicates that ${\bf a}\!\!=\!\![a_1,a_2,...,a_K]^T\!\in\!\mathbb{C}^{K\!\times\!1}$ is a sparse vector and $K_a\!\!\!=\parallel\!\!{\bf a}\!\!\parallel_0\ll\!\!\!K$. Moreover, this activity pattern exhibits the block-sparsity, since ${\bf a}$ typically remains unchanged in $J$ successive time slots within a frame \cite{BWang1,YangDU1,Profshim2,TWOLEVEL}. Furthermore, ${\bf x}_k^j\!=\!a_k g_k^j {\bf d}_k^j$ for $\forall j\!\!\in\!\![J]$ exhibits the structured sparsity \cite{MBMMUD1,MBMMUD2}, due to the sparse nature of media modulated symbols' feature as illustrated in (\ref{eq:d}). The block-sparsity and structured sparsity of the UL signals inspire us to invoke CS theory to detect the active devices and demodulate the data at the BS.

To exploit the block-sparsity of active MTD patterns, we first rewrite the received signals within a frame as
\vspace{-1mm}
\begin{equation}\label{eq:systemModel}
\begin{array}{l}
\bf{Y}=\bf{ H}\bf{X}+\bf{W},
\end{array}
\end{equation}
where we have ${\bf Y}\!=\![{\bf y}^1, {\bf y}^2, ..., {\bf y}^J]\in\mathbb{C}^{N_r \times J}$, ${{\bf H}}\in\mathbb{C}^{N_r \times (K N_t)}$, ${\bf X}\!=\![\widetilde{\bf x}^1, \widetilde{\bf x}^2, ..., \widetilde{\bf x}^J]\in\mathbb{C}^{(K N_t) \times J}$, and ${\bf W}=[{\bf w}^1, {\bf w}^2, ..., {\bf w}^J]\in\mathbb{C}^{N_r \times J}$.
Thus the massive access problem can be formulated as the following optimization problem
\begin{equation}\label{eq:OPTproblem}
\begin{split}
&\min\nolimits_{\bf X} \parallel {\bf Y}-{\bf HX}\parallel_F^2=\min\nolimits_{\{{\bf \widetilde x}^{j}\}_{j=1}^{J}}\sum\nolimits_{j=1}^J\parallel {\bf y}^j-{\bf  H}{\bf \widetilde x}^j\parallel_2^2\\
&=\min\nolimits_{\{a_k,{\bf d}_k^j,g_k^j\}_{j=1,k=1}^{J,K}} \,\, \sum\nolimits_{j=1}^J\parallel {\bf y}^j-\sum\nolimits_{k = 1}^K a_k g_k^j {\bf H}_k{\bf d}_k^j\parallel_2^2\\
&~{\rm s.t.}~~ (2)~~{\rm and}~~\parallel\!\!{\bf a}\!\!\parallel_0\ll K.
\end{split}
\end{equation}

\SetAlFnt{\scriptsize}
\SetAlCapFnt{\normalsize}
\SetAlCapNameFnt{\normalsize}
\begin{algorithm}[tp!]
\caption{{\color{black}Proposed StrOMP Algorithm}}
\label{Algorithm:1}
\KwIn{${\bf Y}\in\mathbb{C}^{N_r \times J}$, ${\bf H}\in\mathbb{C}^{N_r \times (K N_t)}$, and threshold $P_{\rm th}$.}
\KwOut{The index set of estimated active MTDs ${\Gamma}\subseteq [K]$, $\widehat{K_a}=|\Gamma|_c$.}
{\bf Initialization}: The iterative index $i$=1, the residual matrix ${\bf R}^{(0)}\!\!=\!\!{\bf Y}$, $\Gamma^{(0)}\!\!=\!\!\emptyset$. We define ${\bf m}\!\!\in\!\!\mathbb{C}^{K\!\times\!1}$ as an intermediate block correlation variable. For possible active MTDs given their temporary index set $\Lambda$, their MAP's index set is $\widetilde\Lambda\!\!=\!\!\{\widetilde \Lambda_n\}_{n=1}^{|\Lambda|_c}$, where $\widetilde\Lambda_n\!\!=\!\!\{N_t(\Lambda[n]\!\!-\!\!1)\!+\!u\}_{u=1}^{N_t}$ is the MAP's index set of the $n$-th MTD in $\Lambda$ for $ n\in[|\Lambda|_c]$\;
\While{$1$}{
$[{\bf m}]_k=\sum\nolimits_{l=(k-1)N_t+1}^{kN_t}\sum\nolimits_{j=1}^J|({\bf H}_{[:,l]})^H{\bf R}_{[:,j]}^{(i-1)}|^2,~{\rm for}~k\in [K]$\;
\label{block sparsity}
$k^{\star}={\rm arg\mathop{max}\nolimits}_{\widehat{k}\in[K]}{[{\bf m}]_{\widehat k}}$\;
\label{max}
$\Lambda=\Gamma^{(i-1)}\cup k^{\star}$;~~~\{Possible support estimate\}\\
\label{support estimate}
${\bf B}_{[\widetilde\Lambda,:]}\!=\!({\bf H}_{[:,\widetilde\Lambda]})^\dagger{\bf Y},{\bf B}_{[[KN_t]\setminus \widetilde\Lambda, :]}=0$;\{Coarse signal estimate via LS\}\\
\label{LS1}
\color{black}{
$\eta^{\star}_{n,j}={\rm arg\mathop{max}\nolimits}_{\widehat{\eta}_{{n,j}}\in\widetilde\Lambda_{n}}|{\bf B}_{[{\widehat{\eta}_{n,j}},j]}|^2,~{\rm for}~n\in[|\Lambda|_c],~j\in [J]$\;
\label{structure1}
$\Omega^{(j)}=\{\eta^{\star}_{n,j}\}_{n=1}^{|\Lambda|_c},~{\rm for}~j\in [J]$\;
${\bf A}_{[\Omega^{(j)},j]}\!\!=\!\!({\bf H}_{[ :,\Omega^{(j)}]})^\dagger{\bf Y}_{[:,j]},{\bf A}_{[[KN_t]\!\setminus\!{\Omega^{(j)}}, j]}=0,~{\rm for}~j\!\!\in\!\! [J]$;~~\{Fine signal estimate via LS\}\\
\label{LS2}
${\bf R}^{(i)}={\bf Y}-{\bf H}{\bf A}$;~~~\{Residue Update\}\\
\label{Residual Update}
\uIf {$\left\|{\bf R}^{(i-1)}\right\|_F - \left\|{\bf R}^{(i)}\right\|_F < P_{\rm th}$}{
\label{stop cretiria}
{\bf break};~~~\{Terminates the while-loop\}\\
}\Else{
$\Gamma^{(i)}=\Lambda$;~~~\{Support estimate update\}\\
\label{supportupdate}
$i=i+1$\;
\label{iterationupdate}
}
}
}
\KwResult{$\Gamma=\Gamma^{(i-1)}$ , ~$\widehat{K_a}=|\Gamma|_c$.}
\label{Algorithm end 1}
\end{algorithm}

In the following subsections, we will first utilize the proposed StrOMP algorithm to determine the indices of active devices. On that basis, the associated data is further detected based on the proposed SIC-SSP algorithm. {\color{black} Finally, we will discuss the computational complexity of the proposed algorithms.}

\SetAlFnt{\scriptsize}
\SetAlCapFnt{\normalsize}
\SetAlCapNameFnt{\normalsize}
\begin{algorithm}[tp!]
\caption{Proposed SIC-SSP Algorithm}
\label{Algorithm:2}
\KwIn{${\bf Y}\!=\![{\bf y}^1, {\bf y}^2, ..., {\bf y}^J]\in\mathbb{C}^{N_r \times J}$, ${\bf H}\in\mathbb{C}^{N_r \times (K N_t)}$, and the output of Algorithm 1: $\Gamma$, $\widehat{K_a}$.}
\KwOut{Reconstructed UL access signal ${\bf X}\!=\![\widetilde{\bf x}^1, \widetilde{\bf x}^2, ...,\widetilde{\bf x}^J]$.}
\For{$j={1}:J$}{
\label{outerloop}
\For{$s={1}:\widehat{K_a}$}{
\label{innerloop}
\If {$s=1$}{
${\bf v}={\bf y}^j$,~$\Lambda=\Gamma$, where ${\bf v}$ is the measurement vector and $\Lambda$ is the remaining set of MTDs to be decoded, and the definitions of $\widetilde{\Lambda}$ and $\widetilde{\Lambda}_n$ are the same as those in Algorithm 1;\{Initialization\}\\
}
$i=1,$~$\Psi^{(0)}=\emptyset$,~${\bf r}^{(0)}={\bf v}$;~~~\{Initialization\}\\
\While{$1$}{
\label{while}
$[{\bf p}]_{\widetilde\Lambda}=({\bf H}_{[:,\widetilde\Lambda]})^H{\bf r}^{(i-1)}$, $[{\bf p}]_{[KN_t]\setminus \widetilde\Lambda}=0$;~~~\{Correlation\}\\
$\tau^{\star}_n={\rm arg\mathop{max}\nolimits}_{\widehat{\tau}_n\in \widetilde{\Lambda}_n}|[{\bf p}]_{\widehat{\tau}_n}|^2,~{\rm for}~n\in [|\Lambda|_c]$\;
\label{structuredSpar1}
$\Omega\!=\!\{\tau^{\star}_n+(\Lambda[n]-1)N_t\}_{n=1}^{|\Lambda|_c}$;\{$|\Lambda|_c$ most likely MAPs\}\\
$\Omega'=\Omega\cup\Psi^{(i-1)}$;\{Preliminary support estimate\}\\
$[{\bf e}]_{\Omega'}\!=\!({\bf H}_{[:,\Omega']})^\dagger{\bf r}^{(0)},[{\bf e}]_{[KN_t]\setminus \Omega'}=0$;\{Coarse LS\}\\
$\eta^{\star}_n={\rm arg\mathop{max}\nolimits}_{\widehat{\eta}_n\in \widetilde{\Lambda}_n}|[{\bf e}]_{\widehat{\eta}_n}|^2,~{\rm for}~n\in [|\Lambda|_c]$\;
\label{structuredSpar2}
$\Psi^{(i)}=\{\eta^{\star}_n+(\Lambda[n]-1)N_t\}_{n=1}^{|\Lambda|_c}$;~~~\{Pruning support set\}\\
$[{\bf e}]_{\Psi^{(i)}}\!=\!({\bf H}_{[:,\Psi^{(i)}]})^\dagger{\bf r}^{(0)}\!,\![{\bf e}]_{[KN_t]\setminus \Psi^{(i)}}\!=\!0$;\{Fine LS\}\\
\label{estimation}
${\bf r}^{(i)}={\bf r}^{(0)}-{\bf H}{\bf e}$;\{Residue Update\}\\
\If {$i\geq \widehat{K_a}$~~{\rm or}~~$\Psi^{(i)}=\Psi^{(i-1)}$}{
\label{startSIC}
$\Psi=\Psi^{(i)}$, $n^\star={\rm arg\mathop{max}\nolimits}_{\widehat n\in [|\Lambda|_c]}|[{\bf e}]_{\Psi[\widehat{n}]}|^2$;\\
\label{SIC1}
${\bf v}\!=\!{\bf v}\!-\!{\bf H}_{[:,\Psi[n^\star]]}[{\bf e}]_{\Psi[n^\star]}$;\{Measurement vector update\}\\
\label{SIC2}
$[\widetilde{\bf x}^j]_{\Psi[n^\star]}=[{\bf e}]_{\Psi[n^\star]}$,~~~$\Lambda=\Lambda\setminus\{\Lambda[n^\star]\}$;\\
\label{SIC3}
${\bf break}$;~~~\{Terminates the while-loop\}\\
\label{break}
}
\label{endSIC}
$i=i+1$\;
}\label{endwhile}
}\label{innnerloopend}
}\label{outterloopend}
\KwResult{${\bf X}\!=\![\widetilde{\bf x}^1, \widetilde{\bf x}^2, ...,\widetilde{\bf x}^J]$.}
\label{Algorithm end 2}
\end{algorithm}

\vspace{-3mm}
{\color{black}
\subsection{ Proposed StrOMP Algorithm for AUD}
}

{\color{black}The proposed StrOMP algorithm listed in Algorithm \ref{Algorithm:1}, is developed from the orthogonal matching pursuit (OMP) algorithm of \cite{OMP}. Specifically, line \ref{block sparsity} calculates the sum correlation ${\bf m}$ associated with all $N_t$ MAPs in $J$ time slots for each MTD; line \ref{support estimate} combines $k^{\star}$ (i.e., the most likely active MTD) with $\Gamma^{(i-1)}$ to update the possible support set $\Lambda$; in line \ref{LS1}, the coarse signal estimate is obtained by the least squares (LS) algorithm; lines \ref{structure1}$\sim$\ref{LS2} exploit the structured sparsity of media modulated symbols to estimate the possible MAPs based on the coarsely estimated signal ${\bf B}$, and then the fine signal estimate is obtained in line \ref{LS2} for improved robustness to noise; line \ref{Residual Update} updates the residual by using the finely estimated signal ${\bf A}$. In line \ref{stop cretiria}, if the energy difference of the residual in adjacent iterations $\left\|{\bf R}^{(i-1)}\right\|_F - \left\|{\bf R}^{(i)}\right\|_F$ falls below a predefined threshold, the loop stops, otherwise the iteration continues.}

{\color{black}The classical OMP algorithm requires the sparsity level $K_a$, whereas the proposed StrOMP algorithm adaptively acquires the number of active MTDs without knowing $K_a$. Compared to the OMP algorithm, the proposed StrOMP achieves an improved detection performance by exploiting the block-sparsity (line \ref{block sparsity}) and the structured sparsity (lines \ref{structure1}$\sim$\ref{LS2}) of the UL access signals.}

\begin{figure*}[!t]
\centering
\subfigure{
    \begin{minipage}[t]{0.5\linewidth}
        \centering
\label{fig:PeSNR}
        \includegraphics[width=3.2in]{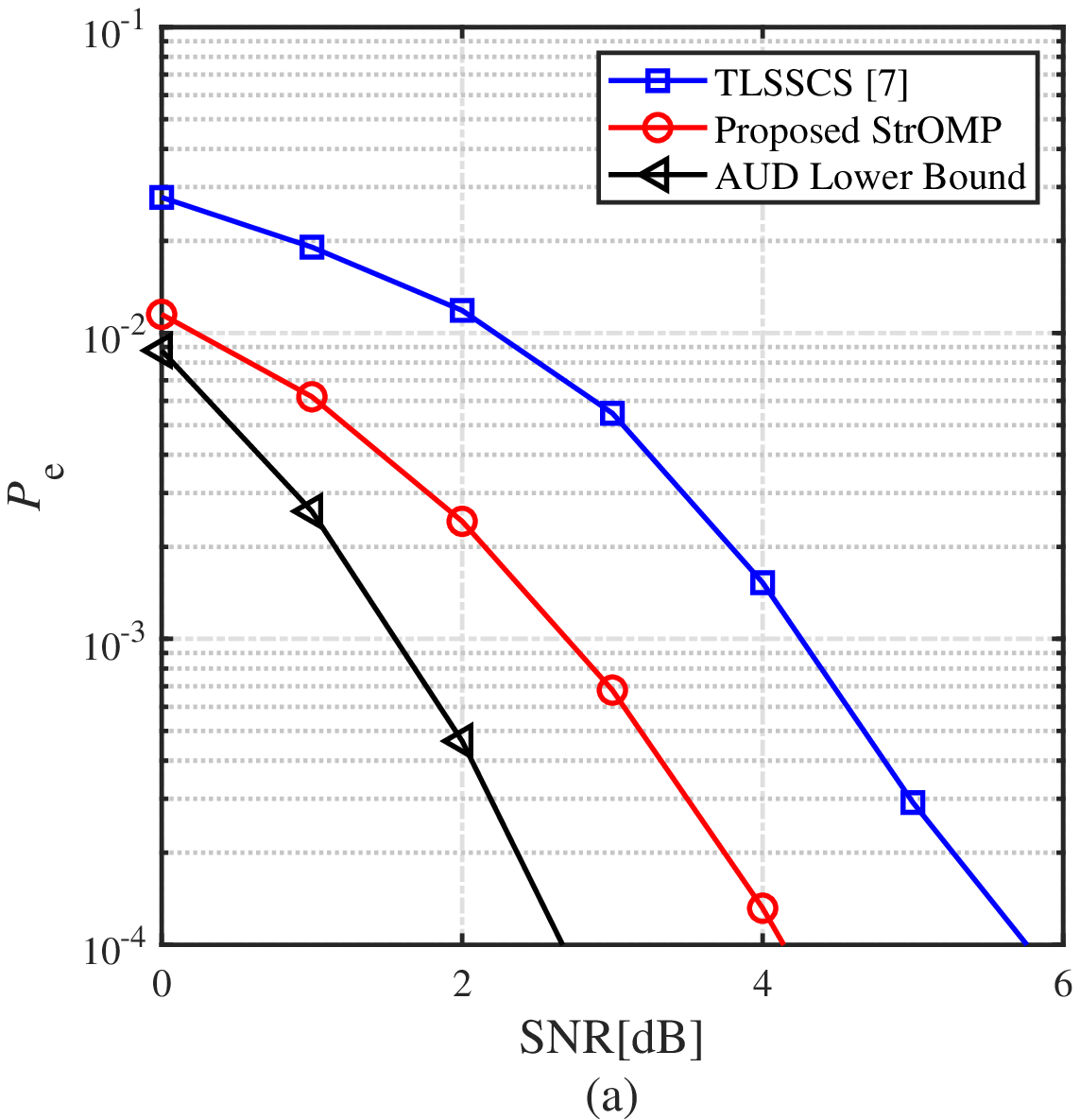}\\
        \vspace{0.02cm}
    \end{minipage}%
}%
\subfigure{
    \begin{minipage}[t]{0.5\linewidth}
        \centering
\label{fig:BERSNR}
        \includegraphics[width=3.2in]{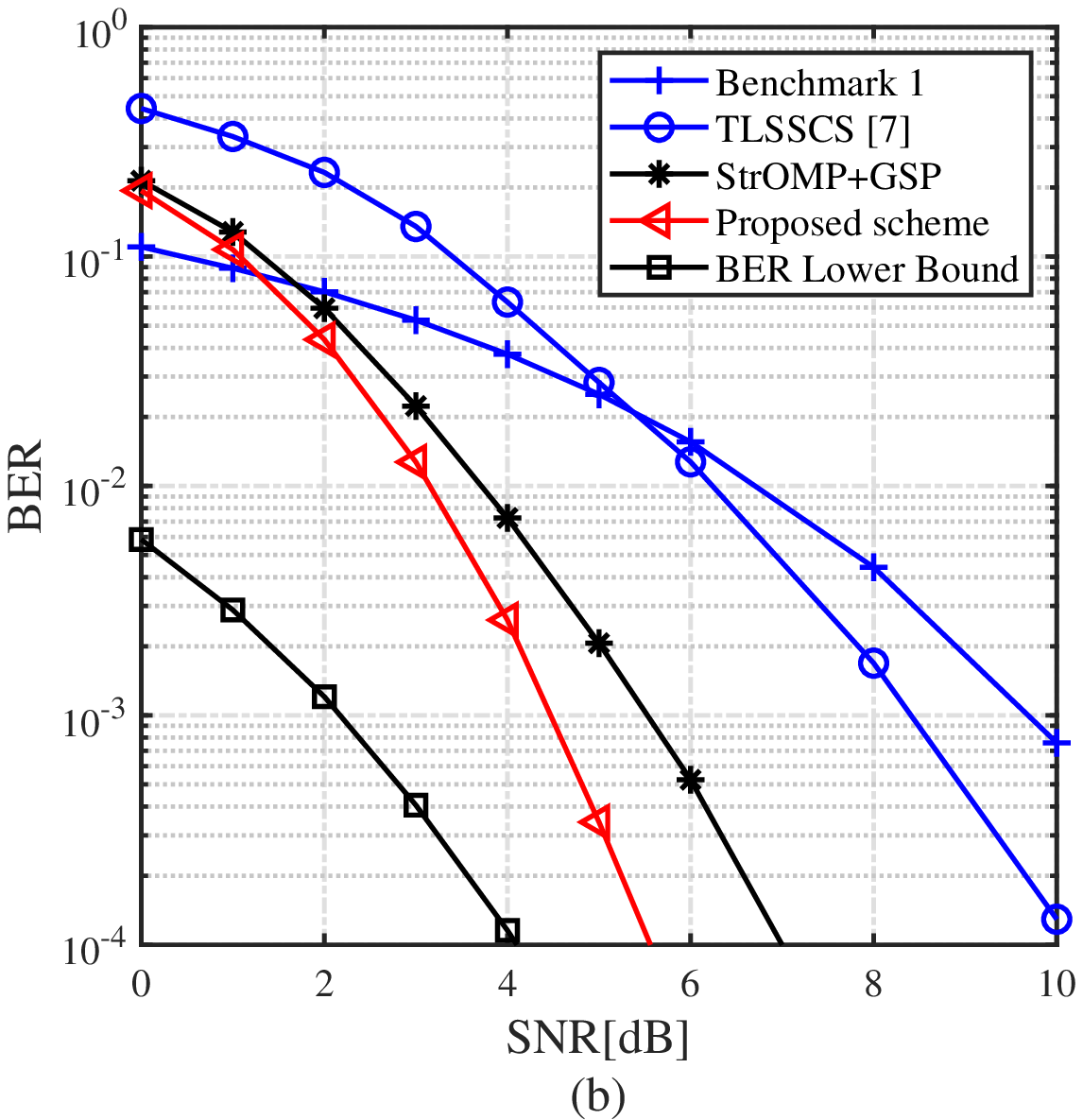}\\
        \vspace{0.02cm}
    \end{minipage}%
}%
\centering
\setlength{\abovecaptionskip}{-1.8mm}
\captionsetup{font={footnotesize, color = {black}}, name={Fig.},labelsep=period}
\caption{Performance comparison of different solutions versus the SNR ($N_r=50$, $J=12$): (a) AUD performance; (b) BER performance.}
\label{fig:PerforSNR}
\end{figure*}

\subsection{The SIC-SSP Algorithm Proposed for Data Detection}

Based on the estimated active MTDs $\Gamma$ obtained from Algorithm 1, the data detection problem in formula (\ref{eq:OPTproblem}) reduces to the same CS problem as in \cite{Gao} (i.e., Eq. (10) for $J=1$ in \cite{Gao}), which can be solved by the group subspace pursuit (GSP) algorithm of \cite{Gao}. To further improve the performance, the proposed SIC-SSP algorithm, as listed in Algorithm \ref{Algorithm:2}, intrinsically integrates the idea of successive interference cancellation (SIC) with the GSP algorithm. Specifically, the outer for-loop recovers $\{\widetilde{\bf x}^j\}_{j=1}^{J}$ separately. For each $\widetilde{\bf x}^j$ with $j\!\!\in\!\![J]$, the inner for-loop recovers a structured sparse signal with $\widehat{K_a}$ sparsity by performing ($\widehat{K_a}-1$) SIC operations. In contrast to the existing GSP algorithm, the inner for-loop of the proposed algorithm incorporates the SIC operation (line \ref{startSIC}$\sim$\ref{endSIC}). Specifically, line \ref{SIC1} selects the index of the maximum element of the finely estimated signal ${\bf e}$ and subsequently line \ref{SIC2} eliminates it from the measurement vector ${\bf v}$; line \ref{SIC3} records the maximum element in $\widetilde{\bf x}^j$ ($j\!\!\in\!\![J]$) and reduces the size of the remaining set of active MTDs $\Lambda$ by 1, which corresponds to reducing the column dimension of the channel matrix in the next iteration for improving the data detection performance. Moreover, lines \ref{structuredSpar1} and \ref{structuredSpar2} improve the performance by exploiting the signal's structured sparsity. Finally, the algorithm is terminated when ${\bf X}$ is fully reconstructed.

\vspace{-3mm}
\color{black}\subsection{Computational Complexity}
\begin{enumerate}[]
\item {The computational complexity of the proposed StrOMP algorithm (Algorithm 1) in the $i$-th iteration mainly depends on the following operations.

{\bf Signal correlation} (line \ref{block sparsity}): The matrix multiplication involved has the complexity on the order of $\mathcal{O}(JKN_tN_r)$.

{\bf Coarse signal estimate via LS} (line \ref{LS1}): Coarse LS solution has the computational complexity on the order of $\mathcal{O}(J(2N_r(iN_t)^2+(iN_t)^3))$.

{\bf Fine signal estimate via LS} (line \ref{LS2}): Fine LS solution has the computational complexity on the order of $\mathcal{O}(J(2N_ri^2+i^3))$.

{\bf Residue update} (line \ref{Residual Update}): Since signal ${\bf A}$ acquired in line \ref{LS2} is a sparse matrix, the complexity of computing the residual is $\mathcal{O}(JN_ri)$.}

\item {The computational complexity of the proposed SIC-SSP algorithm (Algorithm 2) in the $s$-th ($1\leq s \leq \widehat{K_a}$) inner for-loop mainly depends on the following operations.

{\bf Correlation} (line 8): The matrix multiplication involved has the complexity on the order of $\mathcal{O}((\widehat{K_a}-s+1)N_tN_r)$.

{\bf Coarse LS} (line 12): Coarse LS solution has the computational complexity on the order of $\mathcal{O}(2N_r(2(\widehat{K_a}-s+1))^2+(2(\widehat{K_a}-s+1))^3)$.

{\bf Fine LS} (line 15): Fine LS solution has the computational complexity on the order of $\mathcal{O}(2N_r(\widehat{K_a}-s+1)^2+(\widehat{K_a}-s+1)^3)$.

{\bf Residue update} (line 16): The complexity of computing the residual is $\mathcal{O}((\widehat{K_a}-s+1)N_r)$.}
\end{enumerate}

\begin{figure*}[!t]
\centering
\subfigure{
    \begin{minipage}[t]{0.5\linewidth}
        \centering
\label{fig:PeT}
        \includegraphics[width=3.2in]{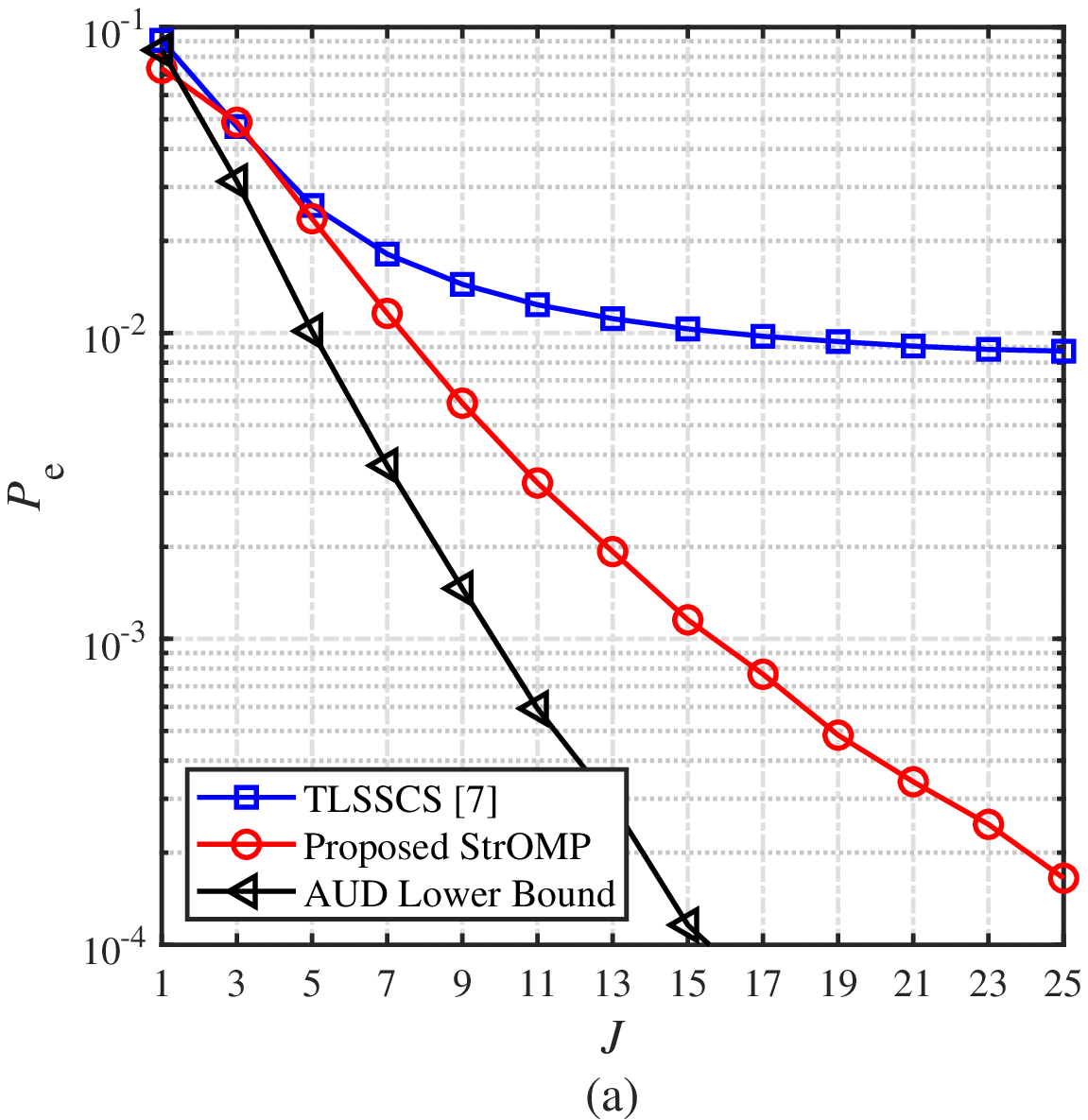}\\
        \vspace{0.02cm}
    \end{minipage}%
}%
\subfigure{
    \begin{minipage}[t]{0.5\linewidth}
        \centering
\label{fig:BERT}
        \includegraphics[width=3.2in]{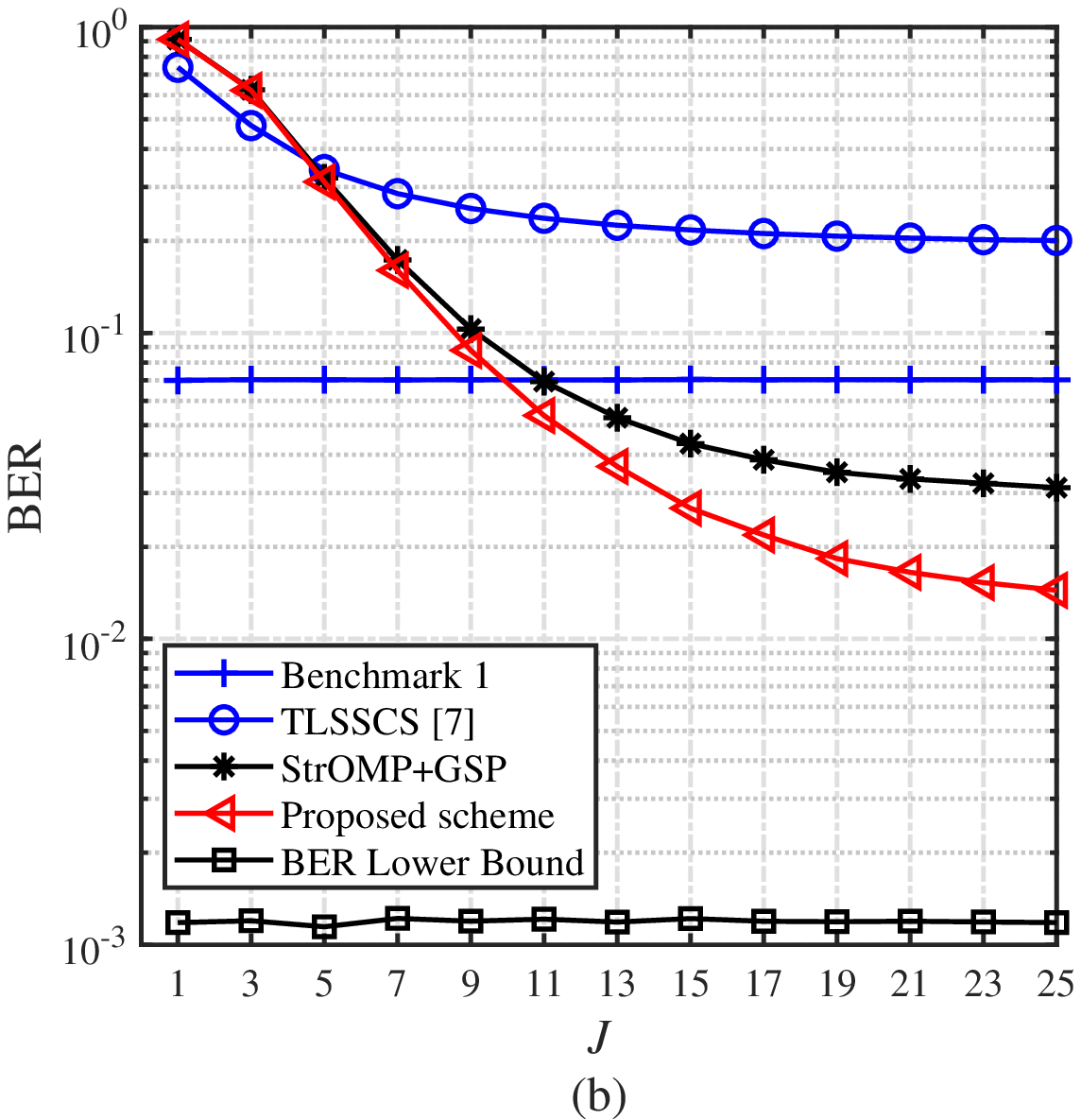}\\
        \vspace{0.02cm}
    \end{minipage}%
}%
\centering
\setlength{\abovecaptionskip}{-1.8mm}
\captionsetup{font={footnotesize, color = {black}}, name={Fig.},labelsep=period}
\caption{Performance comparison of different solutions versus the frame length $J$ (SNR~=~2 dB, $N_r=50$): (a) AUD performance; (b) BER performance.}
\label{fig:PerforT}
\end{figure*}

\begin{figure*}[!t]
\centering
\subfigure{
    \begin{minipage}[t]{0.5\linewidth}
        \centering
\label{fig:PeNr}
        \includegraphics[width=3.2in]{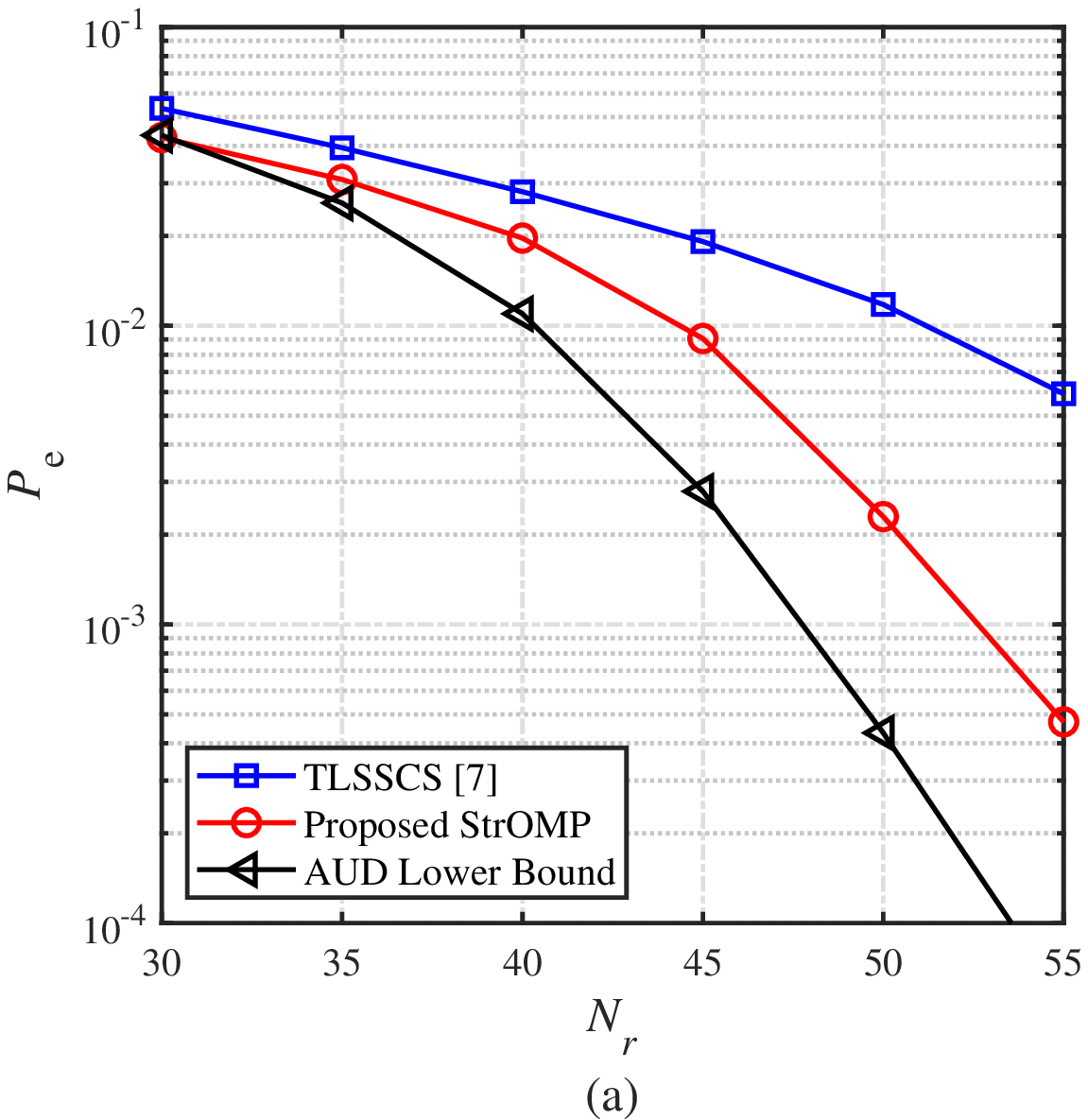}\\
        \vspace{0.02cm}
    \end{minipage}%
}%
\subfigure{
    \begin{minipage}[t]{0.5\linewidth}
        \centering
\label{fig:BERNr}
        \includegraphics[width=3.2in]{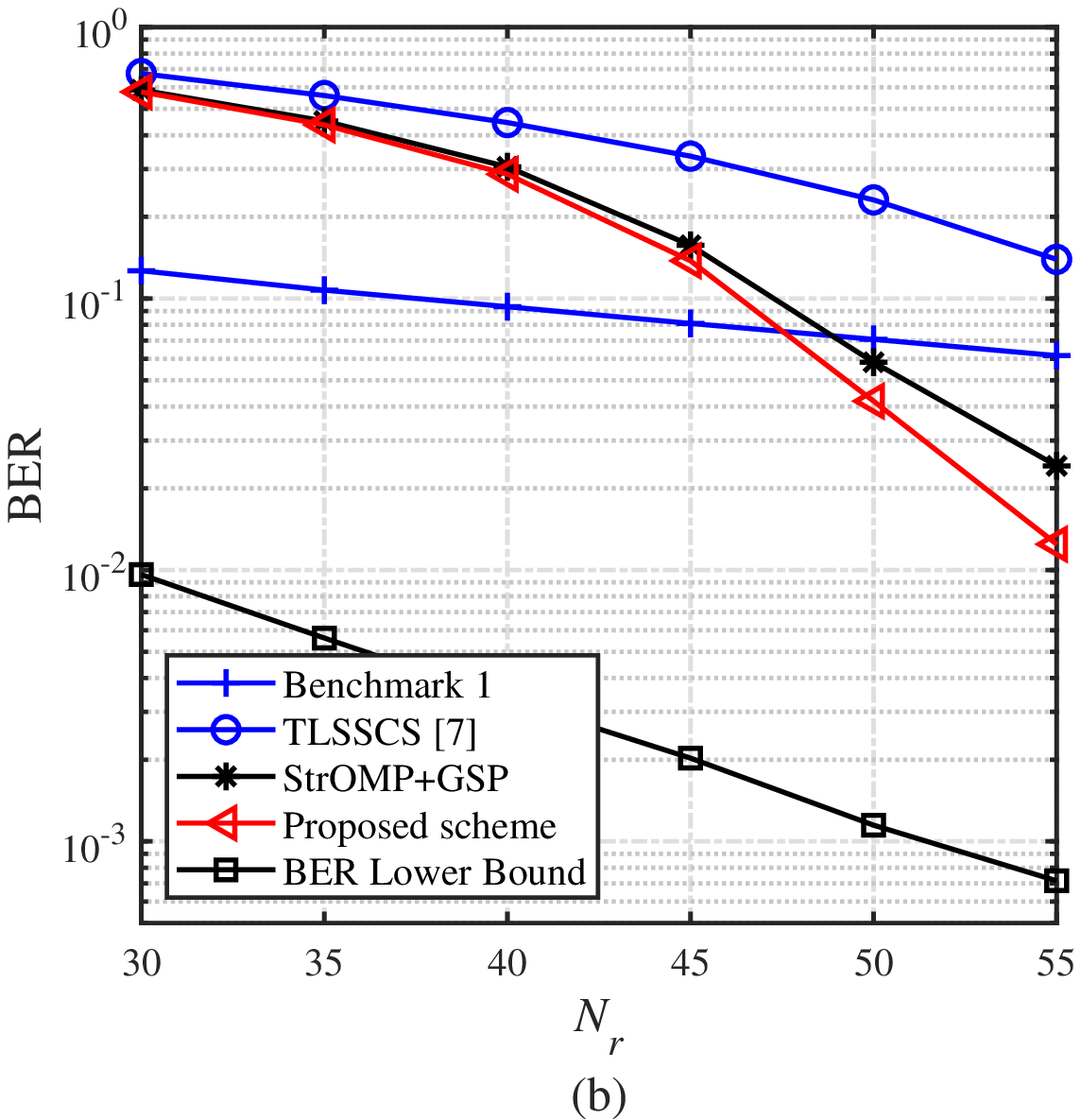}\\
        \vspace{0.02cm}
    \end{minipage}%
}%
\centering
\setlength{\abovecaptionskip}{-1.8mm}
\captionsetup{font={footnotesize, color = {black}}, name={Fig.},labelsep=period}
\caption{Performance comparison of different solutions versus the the number of receive antennas $N_r$ (SNR~=~2 dB, $J=12$): (a) AUD performance; (b) BER performance.}
\label{fig:PerforNr}
\end{figure*}

\color{black}\section{Simulation Results}

Let us now evaluate {\color{black}the probability of AUD error rate ($P_{\rm e}$) and} the bit error rate (BER) for the proposed CS-based massive access solution. Here {\color{black}$P_{\rm e}=\frac{E_u+E_f}{K}$}, and ${\rm BER}\!=\!\frac{E_u J\eta+B_m+B_c}{K_a J\eta}$, where $E_u$ is the number of active MTDs missed by activity detection, {\color{black}$E_f$ is the number of falsely detected inactive MTDs}, $B_m$ and $B_c$ are the total number of error bits in the media modulated symbols and conventional symbols for detected active MTDs within a frame, respectively, and $K_a J\eta$ is the total number of bits transmitted by $K_a$ active MTDs within a frame. In our simulations, the total number of MTDs is $K=100$ with $K_a=8$ active MTDs. Furthermore, each media modulation based MTD adopts $M_r=2$ RF mirrors and 4-QAM ($M=4$), hence the overall throughput becomes $\eta=M_r+{\rm log}_2M=4$ bpcu. {\color{black}Finally, $P_{\rm th}$ in the proposed StrOMP algorithm is set to 2, which is selected experimentally.}

For comparison, we consider the following benchmarks. {\bf Benchmark 1}: Zero forcing multi-user detector for the traditional mMIMO UL \cite{Gao} with $K_a$ single-antenna users adopting 16-QAM to achieve the same 4 bpcu. {\bf TLSSCS}: TLSSCS detector from literature \cite{TWOLEVEL}, and the scaling factor $\alpha=4$ (i.e., $\alpha$ in Eq. (6) of \cite{TWOLEVEL}). {\color{black}{\bf StrOMP+GSP}: The proposed StrOMP algorithm and the existing GSP algorithm of \cite{Gao} are successively used to detect the active MTDs and the data. {\bf AUD lower bound}: A modified StrOMP algorithm relying on the perfect knowledge of $K_a$, which performs the iterations including lines \ref{block sparsity}$\sim$\ref{Residual Update} and lines \ref{supportupdate}$\sim$\ref{iterationupdate} for $K_a$ times, and the output estimated support set is $\Gamma^{(K_a)}$ containing $K_a$ elements.} {\bf BER lower bound}: The Oracle LS based detector relying on the perfect known index set of active MTDs and the support set of media modulated symbols, is considered as the BER lower bound of the proposed mMTC scheme.

{\color{black}
From Fig. \ref{fig:BERSNR}, Fig. \ref{fig:BERT}, and Fig. \ref{fig:BERNr}, it is obvious that the BER performance of the proposed mMTC scheme outperforms the traditional mMIMO UL (Benchmark 1) for the same throughput when $P_{\rm e}$ is small enough, thanks to the extra bits introduced by media modulation. Note that it is actually unfair for the proposed scheme to be compared with the benchmark 1 in BER performance since the latter does not consider the AUD error.
}

{\color{black}
Fig. \ref{fig:PeSNR} and Fig. \ref{fig:BERSNR} compare the AUD performance and BER performance versus the signal-to-noise ratio (SNR), respectively. It is clear that the AUD performance of the proposed StrOMP algorithm is better than the TLSSCS algorithm, and closer to the AUD lower bound. We find that the BER performance of our ``StrOMP+SIC-SSP" solution outperforms the TLSSCS detector, and the ``StrOMP+GSP" solution, which demonstrates the efficiency of the proposed solution. Moreover, compared to the ``StrOMP+GSP" solution, the BER performance of our ``StrOMP+SIC-SSP" solution is getting better and better with the increase of SNR, which proves the the efficiency of the SIC operation.
}

\begin{table*}[!t]
\centering
\captionsetup{font = {normalsize, color = {black}}, labelsep = period} 
\color{black}\caption*{Table II: Computational complexity comparison of different algorithms}
\begin{threeparttable}
\begin{tabular}{|p{2cm}|p{3cm}|p{6cm}|p{2cm}|p{2cm}|}
\Xhline{1.2pt}
\multicolumn{2}{|c|}{\multirow{2}*{{\bf Algorithms}}} & \multirow{2}*{{\bf Computational complexity}}& \multicolumn{2}{|c|}{{\bf Complex-valued multiplications\tnote{1} ($10^6$)}}  \\%
\cline{4-5}
\multicolumn{2}{|c|}{~} & ~& $N_r=50$ &$N_r=100$  \\%
\Xhline{1.2pt}
\multirow{3}*{AUD}
~&Proposed StrOMP &$\mathcal{O}((K_a+1)JKN_tN_r+\sum\nolimits_{s=1}^{K_a+1}[JN_r(s+2s^2+2(sN_t)^2)+J(s^3+(sN_t)^3)])$ & 9.6 & 17.6\\
\cline{2-5}
~&AUD part of TLSSCS [7] & $\mathcal{O}((K_a+1)[{N_r}^2(KN_t+J)+N_rJKN_t]+\sum\nolimits_{s=1}^{K_a+1}[{N_r}^2+2N_r(sN_t)^2+(sN_t)^3])$& 12.5 & 44.2\\
\cline{2-5}
~& AUD lower bound & $\mathcal{O}(K_aJKN_tN_r+\sum\nolimits_{s=1}^{K_a}[JN_r(s+2s^2+2(sN_t)^2)+J(s^3+(sN_t)^3)])$&7.1 & 13.2\\
\Xhline{1.2pt}
\multirow{5}*{Data detection}
~&Proposed SIC-SSP&$\mathcal{O}(J\sum\nolimits_{s=1}^{K_a}[2sN_r(N_t+1)+14N_rs^2+11s^3])$&2.1 &4.0\\
\cline{2-5}
~&Data detection part of TLSSCS [7] &$\mathcal{O}(JN_rK_aN_t+2N_r(K_aN_t)^2+(K_aN_t)^3)$ &0.15 & 0.28\\
\cline{2-5}
~& GSP [8] & $\mathcal{O}(J[2sN_r(N_t+1)+14N_r{K_a}^2+11{K_a}^3])$ &0.65 &1.2\\
\cline{2-5}
~& BER lower bound & $\mathcal{O}(JN_rK_a+2N_r{K_a}^2+{K_a}^3)$ &0.01 &0.02\\
\cline{2-5}
~& Benchmark 1 & $\mathcal{O}(JN_rK_a+2N_r{K_a}^2+{K_a}^3)$&0.01&0.02 \\
\Xhline{1.2pt}
\end{tabular}
\begin{tablenotes}
\footnotesize
\item[1] The number of the complex-valued multiplications is calculated under the parameters $J=12$, $N_t=4$, $K=100$, $K_a=8$.
\end{tablenotes}
\end{threeparttable}
\end{table*}

\vspace{3mm}
{\color{black}
Fig. \ref{fig:PeT} and Fig. \ref{fig:BERT} compare the AUD performance and BER performance versus the frame length $J$, respectively. Owing to the exploitation of the block sparsity, it can be seen that the AUD performance of the proposed StrOMP improves with the increase of $J$. Furthermore, for AUD performance, the advantage of the proposed StrOMP algorithm over the TLSSCS algorithm becomes more obvious with the incerease of $J$. We also find that except for the Oracle LS (BER lower bound), the proposed ``StrOMP+SIC-SSP" solution has the lowest BER floor, for sufficiently large $J$.
}

{\color{black}
Fig. \ref{fig:PeNr} and Fig. \ref{fig:BERNr} compare the AUD performance and BER performance versus the number of receive antennas $N_r$, respectively. Observe from Fig. \ref{fig:PerforNr} that when $N_r$ becomes large, the AUD performance or BER of the proposed ``StrOMP+SIC-SSP" solution is better than that of the TLSSCS detector and the ``StrOMP+GSP" solution, which indicates the superiority of the proposed solution for mMIMO.
}

{\color{black}
The computational complexity of different solutions in our simulation are compared in Table II, where different algorithms are divided into two parts based on their functions (i.e., AUD or data detection). It is obvious that the complex-valued multiplications of the proposed StrOMP algorithm is a little smaller than the AUD part of the TLSSCS algorithm (i.e., lines 1-14 of Algorithm 1 in \cite{TWOLEVEL}) when $N_r=50$. If $N_r$ is doubled, the number of complex-valued multiplications of the proposed StrOMP algorithm increases linearly with $N_r$, whereas the complexity of the AUD part of the TLSSCS algorithm is nearly proportional to the square of $N_r$. Hence, it is clear that our StrOMP algorithm is more suitable for mMIMO with large antenna arrays. Furthermore, after obtaining the active MTDs, the data detection part of the TLSSCS algorithm becomes an LS operation (i.e., line 15 of Algorithm 1 in \cite{TWOLEVEL}) with limited BER performance for media modulated signal. Hence, our proposed SIC-SSP algorithm sacrifices some computational complexity for much better data detection performance.
}
\section{Conclusions}

A media modulation based mMTC UL scheme relying on mMIMO detection at the BS was proposed for achieving reliable massive access with an enhanced throughput. The sparse nature of the mMTC traffic motivated us to propose a CS-based solution. First, an StrOMP algorithm was proposed to detect the active MTDs exhibiting block-sparsity and structured sparsity of the UL signals, which improved the performance. Then, an SIC-SSP algorithm was proposed for detecting the data of the detected MTDs by exploiting the structured sparsity of media modulated symbols for enhancing the performance. {\color{black}Furthermore, we analysed the computational complexity of the proposed algorithms.} Finally, our simulation qualified the benefits of the proposed solution.



\begin{thebibliography}{10}

{\color{black}
\bibitem{overview1}
C. Bockelmann \textit{et al.}, ``Massive machine-type communications in 5G: Physical and MAC-layer solutions," {\em IEEE Commun. Mag.,} vol. 54, no. 9, pp. 59-65, Sept. 2016.
}

\bibitem{Profshim2}
B. K. Jeong, B. Shim and K. B. Lee, ``MAP-based active user and data detection for massive machine-type communications," {\em IEEE
Trans. Veh. Technol.,} vol. 67, no. 9, pp. 8481-8494, Sept. 2018.

\bibitem{BWang1}
B. Wang, L. Dai, T. Mir and Z. Wang, ``Joint user activity and data detection based on structured compressive sensing for NOMA," {\em IEEE Commun. Lett.,} vol. 20, no. 7, pp. 1473-1476, Jul. 2016.

\bibitem{YangDU1}
Y. Du, C. Cheng, B. Dong, Z. chen, X. Wang, J. Fang and S. Li, ``Block-sparsity-based multiuser detection for uplink grant-free NOMA," {\em IEEE Trans. Wireless Commun.,} vol. 17, no. 12, pp. 7894-7909, Dec. 2018.

\bibitem{BWang2}
B. Wang, L. Dai, Y. Zhang, T. Mir and J. Li, ``Dynamic compressive sensing-based multi-user detection for uplink grant-free NOMA," {\em IEEE Commun. Lett.,} vol. 20, no. 11, pp. 2320-2323, Nov. 2016.

\bibitem{YangDU2}
Y. Du, B. Dong, Z. Chen, X. Wang, Z. Liu, P. Gao and S. Li, ``Efficient multi-user detection for uplink grant-free NOMA: Prior-information aided adaptive compressive sensing perspective," {\em IEEE J. Select. Areas Commun.,} vol. 35, no. 12, pp. 2812-2828, Jul. 2017.


\bibitem{TWOLEVEL}
X. Ma, J. Kim, D. Yuan and H. Liu, ``Two-level sparse structure based compressive sensing detector for uplink spatial modulation with massive connectivity," {\em IEEE Commun. Lett.,} vol. 23, no. 9, pp. 1594-1597, Sept. 2019.

\bibitem{Gao}
Z. Gao, L. Dai, Z. Wang, S. Chen and L. Hanzo, ``Compressive-sensing based multiuser detector for the large-scale SM-MIMO uplink," {\em IEEE
Trans. Veh. Technol.,} vol. 65, no. 10, pp. 1860-1865, Feb. 2017.

\bibitem{SM1}
L. Xiao, P. Yang, Y. Xiao, S. Fan, M. Di Renzo, W. Xiang and S. Li, ``Efficient compressive sensing detectors for generalized spatial modulation systems," {\em IEEE Trans. Veh. Technol.,} vol. 66, no. 2, pp. 1284-1298, Feb. 2017.

\bibitem{SM2}
L. Xiao, Y. Xiao, P. Yang, J. Liu, S. Li and W. Xiang, ``Space-time block coded differential spatial modulation," {\em IEEE
Trans. Veh. Technol.,} vol. 66, no. 10, pp. 8821-8834, Oct. 2017.

\bibitem{MBMMUD1}
L. Zhang, M. Zhao and L. Li, ``Low-complexity multi-user detection for MBM in uplink large-scale MIMO systems," {\em IEEE Commun. Lett.,} vol. 22, no. 8, pp. 1568-1571, Aug. 2018.

\bibitem{MBMMUD2}
B. Shamasundar, S. Jacob, L. N. Theagarajan and A. Chockalingam, ``Media-based modulation for the uplink in massive MIMO systems,"  {\em IEEE
Trans. Veh. Technol.,} vol. 67, no. 9, pp. 8169-8183, Sept. 2018.

{\color{black}\bibitem{CSreview}
J. W. Choi, B. Shim, Y. Ding, B. Rao, D. I. Kim, ``Compressed sensing for wireless communications: Useful tips and tricks", {\em Commun. Surveys Tuts.,} vol. 19, no. 3, pp. 1527-1550, 2017.}

\bibitem{MBM1}
A. K. Khandani, ``Media-based modulation: A new approach to wireless transmission," in {\em Proc. IEEEInt. Symp. Inf. Theory}, Jul. 2013, pp. 3050-3054.

{\color{black}
\bibitem{MBM2}
Y. Naresh and A. Chockalingam, ``On media-based modulation using RF mirrors," {\em IEEE Trans. Veh. Tech.,} vol. 66, no. 6, pp. 4967-4983, June 2017.}

{\color{black}
\bibitem{MBM3}
E. Basar, ``Media-based modulation for future wireless systems: A tutorial," {\em IEEE Wireless Commun.,} vol. 26, no. 5, pp. 160-166, Oct. 2019.
}

\bibitem{OMP}
J. A. Tropp and A. C. Gilbert, ``Signal recovery from random measurements via orthogonal matching pursuit," in {\em IEEE Trans. Inform. Theory}, vol. 53, no. 12, pp. 4655-4666, Dec. 2007.

\end{thebibliography}
\end{document}